\documentclass[aps,prb,twocolumn,superscriptaddress,floatfix,longbibliography]{revtex4}
\usepackage{amsfonts}
\usepackage{amssymb}
\usepackage{graphicx}
\usepackage{dcolumn}
\usepackage{bm}
\usepackage{amsmath}
\usepackage[colorlinks,linkcolor=magenta,citecolor=blue,urlcolor=blue]{hyperref}
\usepackage{changes}

\begin{document}

\title{Exact non-Hermitian mobility edges in one-dimensional quasicrystal lattice with exponentially decaying hopping and its dual lattice}

\author{Yanxia Liu}\thanks{These authors contributed equally to this work.}
\affiliation{Beijing National Laboratory for Condensed Matter Physics, Institute of
Physics, Chinese Academy of Sciences, Beijing 100190, China}
\author{Yongjian Wang}\thanks{These authors contributed equally to this work.}
\affiliation{Academy of Mathematics and Systems Science, Chinese Academy of Sciences
Beijing 100190, China}
\affiliation{University of Chinese Academy of Sciences
Beijing 100049, China}
\author{Zuohuan Zheng}
\affiliation{Academy of Mathematics and Systems Science, Chinese Academy of Sciences
Beijing 100190, China}
\affiliation{University of Chinese Academy of Sciences
Beijing 100049, China}
\affiliation{College of Mathematics and Statistics, Hainan Normal University
Haikou, Hainan 571158, China}
\author{Shu Chen}
\email{schen@iphy.ac.cn}
\affiliation{Beijing National Laboratory for Condensed Matter Physics, Institute of
Physics, Chinese Academy of Sciences, Beijing 100190, China}
\affiliation{School of Physical Sciences, University of Chinese Academy of Sciences,
Beijing, 100049, China}
\affiliation{Yangtze River Delta Physics Research Center, Liyang, Jiangsu 213300, China}
\date{\today }

\begin{abstract}
We analytically determine the non-Hermitian mobility edges of a one-dimensional quasiperiodic
lattice model with exponential decaying hopping and complex potentials as well as its dual model, which is just a non-Hermitian generalization
of the Ganeshan-Pixley-Das Sarma model with nonreciprocal nearest-neighboring hopping. The presence of non-Hermitian term  destroys the self-duality symmetry and thus prevents us exploring the localization-delocalization point through looking for self-dual points.  Nevertheless, by applying Avila's global theory, the Lyapunov exponent of the Ganeshan-Pixley-Das Sarma model can be exactly derived, which enables us to get an analytical expression of mobility edge of the non-Hermitian dual model.  Consequently, the mobility edge of the original model is obtained by using the dual transformation, which creates exact mappings between the spectra and wavefunctions of these two models.
\end{abstract}

\maketitle

%%%%%%%%authors%%%%%%%%%%%%%%%

%\thanks{Corresponding author: schen@iphy.ac.cn}

%%%%%%%%authors%%%%%%%%%%%

\section{Introduction}
Anderson localization caused by disorder is an everlasting research topic in condensed matter physics \cite{anderson1958absence}.
Although both random disorder  \cite{abrahams1979scaling,lee1985disordered,evers2008anderson,Thouless72} and quasiperiodic potential
\cite{luschen2018,Aubry1980,Kohmoto1983,Thouless1988,roati2008} can induce Anderson localization, a localization-delocalization transition
is absent in low dimensional random disorder systems. On the other hand, the localization-delocalization- transition can occur even in one-dimensional (1D) quasiperiodic systems.
In comparison with the random disorder systems,  the quasiperiodic systems have their advantages for exploring some exact results
due to the existence of duality relation for the transformation between
real and momentum spaces. A simple but typical example is the Aubry-Andr\'{e}
(AA) model \cite{Aubry1980}, which goes through a
localization transition when the quasiperiodical potential strength
exceeds a transition point i.e., self-duality point.
The studied of the various extensions of AA models reveals more diversified transition styles
\cite{Kohmoto1983,Zhou2013,Cai,DeGottardi,Kohmoto2008,WangYC-review}.
The quasiperiodic lattice models with short-range (long-range) hopping processes \cite%
{biddle2011localization,biddle2010predicted,ganeshan2015nearest,li2016quantum,li2017mobility,li2018mobility,DengX}%
or modified quasiperiodic potentials \cite{sarma1988mobility,sarma1990localization,YCWang2020} can support energy-dependent
mobility edges.

The combination of non-Hermiticity and disorder gives rise to  many new perspectives for the localization phenomena.
Due to throwing off the shackles of the Hermiticity constrain, non-Hermitian random matrices contain much
more abundant symmetry classes according to Bernard-LeClair classification
\cite{BL,HYZhou,CHLiu,Sato} than the corresponding Hermitian
Altland-Zirnbauer classification. In the scheme of random matrix theory, non-Hermitian disorder systems
behave differently in the spectral statistics in comparison with the Hermitian systems \cite{Goldsheid,Markum,Molinari,Chalker}.
Recently, the interplay of non-Hermitian effect and Anderson localization have attracted intensive studies in both random disorder systems \cite{hatano1996localization,hatano1998non,kolesnikov2000localization,Gong,ZhangDW,Hughes, tzortzakakis2019non,HuangYi,ZhangDW2}
and quasiperiodic systems \cite{Yuce,longhi2019metal,jazaeri2001localization,jiang2019interplay,zeng2019topological,longhiPRL,ZengQB,Zeng2020,Liutong,Liu2020,Liu2020L,Cai2020}.
%
%
%Most of the non-Hermitian quasiperiodic models lack the self-dual property. For a time there are no unified methods to obtain exact transition points. Researchers show their special prowess to give the exact transition points by analysis of the competition of the larger hopping and on-site potential\cite{jiang2019interplay}, calculation of the winding number of spectrum\cite{longhiPRL}, and looking for the system with self-dual symmetry\cite{Liutong}.
While most previous works on non-Hermitian quasiperiodical systems focused on the systems with only nearest-neighboring hopping, in this work we study a 1D non-Hermitian quasicrystal lattice with long-range hopping and aim to give an analytical result for the mobility edges. Interplay of long-range hopping and quasiperiodic potential may generate some nontrivial localization properties \cite{biddle2010predicted,biddle2011localization,DengX}.  The Hermitian limit of our model (\ref{Equation01}) can be reduced to the Biddle-Das Sarma model \cite{biddle2010predicted}, which supports an analytical expression of mobility edges determined by a self-duality condition. However, the approach of searching self-duality condition is failed for the non-Hermitian Biddle-Das Sarma model, as the non-Hermitian term destroys the self-duality condition in the whole parameter spaces.

In order to explore the exact non-Hermitian mobility edges, we shall take an alternative method and try to get an analytical expression of Lyapunov exponent. The Lyapunov exponent is an important quantity to characterize the localization properties of disorder systems and was applied to obtain exact transition points for the
non-Hermitian quasiperiodic models with nearest-neighbor hopping \cite{Liu2020L, Liu202009} by using Avila's global theory
\cite{Jitomirskaya1999,Avila2015,Avila2017,Avila2008}. The key of this method is to determine the Lyapunov exponent. However,
the analytical expression of the Lyapunov exponent for the system with long-range hopping is difficult to obtain, which prevent us obtaining the exact mobility edges of model (\ref{Equation01}) directly.
%The model with long-range hopping may be evaluated by dual transform of the model with nearest-neigbor hopping, which supports exact transition points.
To make progress, we shall study the dual model of the long-range hopping model, which is obtained by making a dual transformation to the model (\ref{Equation01}). It is interesting to indicate that the dual model ({\ref{model}) is just a non-Hermitian generalization of the Ganeshan-Pixley-Das Sarma model \cite{ganeshan2015nearest} with nonreciprocal (or asymmetrical) nearest-neighboring hopping.
 The Ganeshan-Pixley-Das Sarma model also supports an analytical expression of mobility edges determined by a self-duality condition \cite{ganeshan2015nearest} and is actually a dual model of Biddle-Das Sarma model \cite{WangYJ2}.  The Lyapunov exponent of the Ganeshan-Pixley-Das Sarma model can be obtained by applying Avila's global theory \cite{WangYJ}, which permits us to obtain an analytical formula  of the mobility edge of the dual model ({\ref{model}). The mobility edge not only splits the extended and localized states but also splits the
real and complex eigenvalues.  Since the dual transformation converts the model (\ref{Equation01}) and its dual model ({\ref{model}) each other, and also construct exact mapping between their eigenvalues and eigenstates,  the mobility edge of the model with long-range hopping can be obtained by the substitution of the parameters from the analytical expression of mobility edge of the dual model.

\begin{figure}[tbp]
\includegraphics[width=0.48\textwidth]{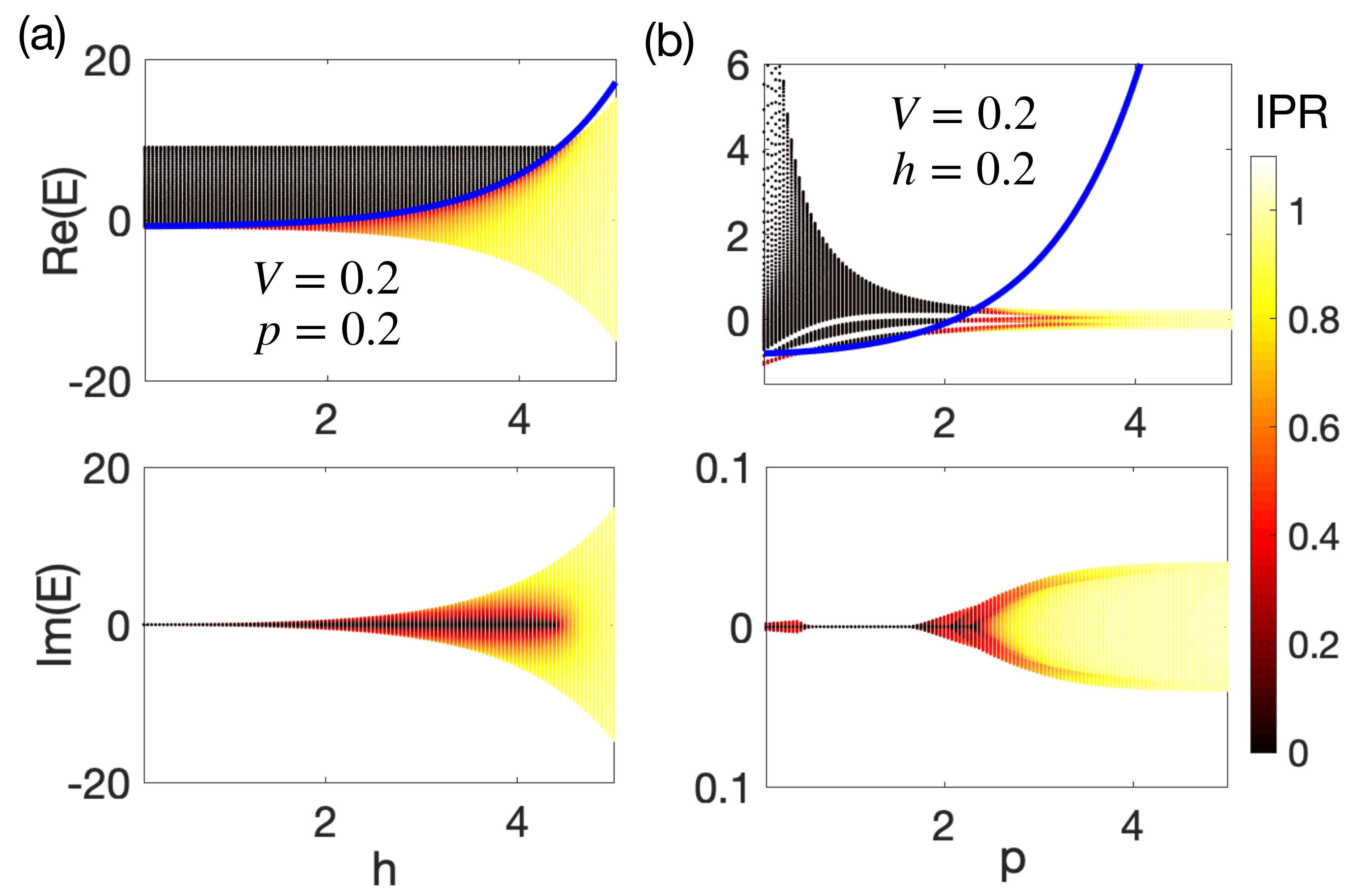}
\caption{(a) The real and imaginary parts of the eigenvalue spectra versus
$h$ for the system with $V =0.2$, $p=0.2$ and $N=610$.
(b) The real and imaginary parts of the eigenvalue spectra versus
$p$ for the system with $V =0.2$, $h=0.2$ and $N=610$.}
\label{fig1}
\end{figure}

\section{Models and results}

%\section{The exact ME of the quasiperiodic model with exponential long-range hopping}

We consider a 1D  quasiperiodic model with long-range hopping terms and a complex
potential, described by
\begin{equation}
E \phi_n=\sum_{n'\neq n} t \mathrm{e}^{-p|n-n'|} \phi_{n'}+V \cos(2\pi \omega n + ih) \phi_n, \label{Equation01}
\end{equation}
where $p>0$ represents the decay rate, $V$ is the quasiperiodic potential strength, $\omega$ is an irrational number,
and $h$ describes a complex phase factor.
We set $t=1$ as the unit of energy  and choose $\omega =\left( \sqrt{5}-1\right) /2$.
When $h=0$ ,i.e., the Hermitian case, the model has been studied in Ref.  \cite{biddle2011localization,biddle2010predicted} and an analytical expression of mobility edge is obtained by applying a self-dual transformation. For the non-Hermitian case with $h\neq 0$, no self-duality relation exists and no analytical result is known. In a recent work \cite{Liu2020}, it has been shown that  there exists mobility edge which can be well fitted by the expression of
$
E=V\mathrm{e}^{|h|}\cosh(p)-1,
$
when both $h$ and $p\gg1$, However, numerical results unveil that the above conjectured expression fails to describe the mobility edge in the region with small $h$ and $g$.

In this work, we analytically derive the exact mobility edge for the model (\ref{Equation01})
with arbitrary $p$ and $h$. The mobility edge segregating the localized and extended states
can be represented as an extremely simple expression:
\begin{equation}
E=V\cosh(p+|h|)-1. \label{MED}
\end{equation}
In general, the eigenvalues of no-Hermitian systems are complex. The above expression indicates that the mobility edge is real, and thus the mobility edge also separate the real and complex states.

Before deriving Eq.(\ref{MED}), we first show the consistency of analytical and numerical results and give numerical verification of
the mobility edge. In order to characterize the localization-delocalization transition of a wavefuntion, we numerically calculate the inverse partition ratio (IPR) of an eigenstate, which
is defined as $\text{IPR}^{(i)}=(\sum_{n}\left\vert \phi_{n}^{i}\right\vert
^{4})/\left( \sum_{n}\left\vert \phi_{n}^{i}\right\vert ^{2}\right) ^{2}$,
where the superscript $i$ labels the $i$th eigenstate of system and $n$
represents the coordinate of lattice site. While IPR$\simeq 1/L$ approaches zero for an extended eigenstate when the lattice size
$L\rightarrow \infty $, IPR$\simeq 1$ for a fully localized eigenstate.
%The transition points (dashed lines) separate localized eigenstates and extended eigenstates in the view of IPR.

In Fig.\ref{fig1}(a), we plot the IPR of different eigenstates versus the complex phase factor $h$ for the system with $V=0.2$ and $p=0.2$.
The eigenstates are characterized by their real and imaginary parts of the corresponding eigenvalues, respectively.
The blue solid line represents the transition points determined by Eq.(\ref{MED}), which separates the extended and localized states.
It is shown that the analytical relation of mobility edge agrees well with numerical results from IPR and spectrum calculations.
With the increase of complex phase factor $h$, the effective complex potential strength increases, and it causes more eigenstates to become localized.
Consequently, the system undergoes the extended, intermediate and localized regime when $h$ increases.
In Fig.\ref{fig1}(b), we plot the real parts and imaginary parts of eigenvalues as well as the IPR of the corresponding
wavefunctions versus the decay rate $p$ by fixing $V=0.2$ and $h=0.2$. The blue solid line marks the transition points determined by Eq.(\ref{MED}), which agree well with
numerical results in the whole region of $p$.

It is quite interesting to indicate that part of eigenstates become localized in the region of a small $p$, As $p \rightarrow 0$, the hopping amplitude between long-range sites decays slowly and long-range hopping terms shall play an important role. Intuitively, this is  counterintuitive because the long-rang hopping generally tends to delocalize the localized states.  To understand why this happens, we consider the limit case with $p=0$ and $V=0$,  in which the model with equal hopping terms has a $(N-1)$-degenerate eigenstates. When a finite $V$ is introduced, $(N-1)$-degenerate eigenstates all become the localized states.
A small $p$ lifts the degeneracy and tends to turn the localized states into the extended states. On the other hand, in the large $p$ limit the hopping amplitude between long-range sites decays very quickly, and the system with a large $p$ can be viewed as a short-rang hopping model with the nearest-neighbor hopping amplitude $t_1=t e^{-p}$ playing a dominate role. If we omitted all the long range terms, the system can be approximately described by the AA model,  and thus all the eigenstates of system become localized as long as $V > 2 t_1$.

Now we return back to discuss how to derive Eq.(\ref{MED}). Instead of directly solving the original model of Eq.(\ref{Equation01}), we shall derive the analytical expression of mobility edge by solving its dual model, i.e., a nearest-neighbor nonreciprocal
hopping non-Hermitian model described by
\begin{equation}
E'u_{j}=\mathrm{e}^{-h}u_{j+1}+\mathrm{e}^{h}u_{j-1}+V_{j}u_{j} \label{model}
\end{equation}
with
\begin{equation}
V_{j}=2\lambda \frac{\cos (2\pi \omega j)}{1-b\cos (2\pi \omega j)},
\end{equation}
where
\begin{equation}
b= S(p)= \frac{1} {\cosh(p)} , %\frac{2\mathrm{e}^{-p}}{1+\mathrm{e}^{-2p}},
\label{parameter1}
\end{equation}
\begin{equation}
\lambda=\frac{S(p)[1-\mathrm{e}^{-p}S(p)]}{V}, \label{lambV}
\end{equation}
and
\begin{equation}
E'=\frac{2 [E+\mathrm{e}^{-p}S(p)]}{V}. \label{energy}
\end{equation}
%with $S(p)=\frac{2\mathrm{e}^{-p}}{1+\mathrm{e}^{-2p}}$.
The model described by Eq.(\ref{model}) can be obtained from model of Eq.(\ref{Equation01}) through the following transformation
\begin{equation}
\phi_n=\sum_{n} \mathrm{e}^{i2\pi\omega n j}u_j.
\end{equation}
Models (\ref{model}) and (\ref{Equation01}) are dual models with parameters connected by Eqs.(\ref{parameter1})-(\ref{energy}).

%The duality is of the essence in dealing with localization transition for the quasicrystal model.
%The eigenstates of one model and its dual model are distinct, and
%the eigenstates of the model are localized, while the eigenstates of its dual model are
%extended, and vice versa. If one model is self-duality, the localization transition point is easy to
%determine. So the self-duality models \cite{biddle2010predicted,biddle2011localization}
%have advantages to determine the MEs.  Although the self-duality of some non-Hermitian system vanishes, the duality
%is always there.

%\section{The exact ME of the dual model}

Now we analytically derive the mobility edge of the model (\ref{model}).
First we consider the case with $h=0$ and calculate the Lyapunov exponent of the system.
The Lyapunov exponent can be evaluated based on the transfer matrix, which can be represented as
\begin{equation}
\gamma \left( E'\right) =\lim_{n\rightarrow \infty }\frac{1}{2\pi n}\int \ln
\left\vert \left\vert T_{n}\left( E',\phi \right) \right\vert \right\vert
\mathrm{d}\phi ,
\end{equation}%
where $||T_n||$ denotes the norm of the transfer matrix given by
\begin{equation*}
T_{n}\left( E',\phi \right) =\prod_{j=1}^{n}M_{j}=\prod_{j=1}^{n}\left(
\begin{array}{cc}
E'-V_{j} & -1 \\
1 & 0%
\end{array}%
\right) .
\end{equation*}%
From the discussions in Ref.\cite{Avila2015,Liu2020L}, we know that
if the energy $E'$ lies in the spectrum of the Hamilton $H$, we have
\begin{equation}
\gamma (E')=\max \{\gamma_c (E'),0\},  \label{LE}
\end{equation}
where $\gamma_c (E')$ is analytically given by
\begin{equation}
\gamma_{c}(E')=\ln |\frac{|bE'+2\lambda|+\sqrt{(bE'+2\lambda)^2-4b^2}}{2(1+\sqrt{1-b^2})}|.  \label{LEh}
\end{equation}

The details for the derivation of the above analytical expression can be found in the appendix A.
While $\gamma \left( E'\right) >0$ corresponds to the localized state,  the extended state is characterized by $\gamma \left( E'\right) =0$.
Therefore the mobility edge can be determined by $\gamma_c (E')=0$ and operator theory, which gives rise to
\begin{equation}
E'=2 \mathrm{sgn} (\lambda)\frac{(1-|\lambda|)}{b}.
\label{ME1}
\end{equation}%
The details for the result of analysis with operator theory can be found in the appendix B.
The mobility edge of the model (\ref{model}) with $h=0$ can also be obtained by looking for the self-dual points of system,
which was originally given in Ref.\cite{ganeshan2015nearest}. Our method based on the analytical expression of Lyapunov exponent results in the same result. Although Eq.(\ref{ME1}) is known, we note that the analytical expression of Lyapunov exponent was only derived recently \cite{WangYJ}.

For the general case with $h \neq 0$,  the system does not have a self-duality point in the parameter space. A nonzero $h$ induces the nonreciprocal hopping, which breaks
the Hermiticity of the system and may cause skin effect for the system with open boundary condition. A similar transformation can transform the non-hermitian
Hamiltonian $H(h)$ under open boundary condition into a hermitian Hamiltonian $H^{\prime }$, via
\begin{equation}
H^{\prime }=SH(h)S^{-1},  \label{smt}
\end{equation}%
where
\[
S=\text{diag} \left( \mathrm{e}^{-h},\mathrm{e}^{-2h},\cdots ,\mathrm{e}^{-Nh}\right)
\]
is a similarity matrix with only diagonal entries and $H^{\prime }=H(h=0)$ is the hermitian Hamiltonian with $h=0$. The
relation between the eigenstates of $%
H$ and $H^{\prime }$ is achieved naturally: $\left\vert \psi \right\rangle
=S^{-1}\left\vert \psi ^{\prime }\right\rangle $. Here  $\left\vert \psi \right\rangle = \sum_j u_j \left\vert j \right\rangle$ is the eigenstate of $H$, and $H \left\vert \psi \right\rangle = E' \left\vert \psi \right\rangle$  gives rise to Eq.(\ref{model}).  The transformation $S^{-1}$ can convert the extended states $%
\left\vert \psi ^{\prime }\right\rangle $ into skin states, which exponentially gather wave function all to one of
boundaries \cite{jiang2019interplay,Yao,Kunst,Alvarez,Xiong,Lee}.

A localized state of $H^{\prime }$ may be expressed in a unified compact form
\[
\left\vert u_{i}\right\vert \propto \mathrm{e}^{- \left\vert i-i_{0}\right\vert/\xi
},
\]
where $i_{0}$ represents the position of localization center of a given localized state, $\xi=1/\gamma$ is
the localization length, and $\gamma$ is
the Lyapunov exponent of the localized state for the system of $h=0$. Then the corresponding wavefunction of
$H(h)$ takes the following form:
\begin{equation}
\left\vert u_{i}\right\vert \propto e^{h i- \gamma \left\vert i-i_{0}\right\vert},
 \label{wavelocal}
\end{equation}
which exhibits different decaying behaviors on different sides of the localization
center.
When $\left\vert h\right\vert \geq \gamma $, delocalization occurs
on one side \cite{jiang2019interplay}, and thus the transition point from the localized state to skin state is given by
\begin{equation}
\gamma =\left\vert h \right\vert . \label{LDP}
\end{equation}
Since a localized state apart from boundaries is not affected by the boundary condition of the
system, it then follows that the boundary of localization-delocalization transition under the
periodic boundary condition is also given by Eq.(\ref{LDP}).
Bringing Eq.(\ref{LE}) into Eq.(\ref{LDP}),  we can get the mobility edges which
can be expressed as
\begin{equation}
E'_c =\frac{2 \text{sgn}(\lambda)\left(\cosh |h|+\sqrt{1-b^2}\sinh |h|-|\lambda| \right)}{b}. \label{LEC}
\end{equation}
The Eq.(\ref{LEC}) with $h=0$ can be reduced to the Eq.(\ref{ME1}).
Replacing the parameters $b$, $\lambda$, and $E'$ with $p$, $V$, and $E$, we can rewrite Eq.(\ref{LEC})
as Eq.(\ref{MED}). Although the mobility edges of a model can be read out from its dual model,
the eigenstates of one model and its dual model are distinct.
If the eigenstates of the model are localized,  the eigenstates of its dual model are extended, and vice versa.

\begin{figure}[tbp]
\includegraphics[width=0.5\textwidth]{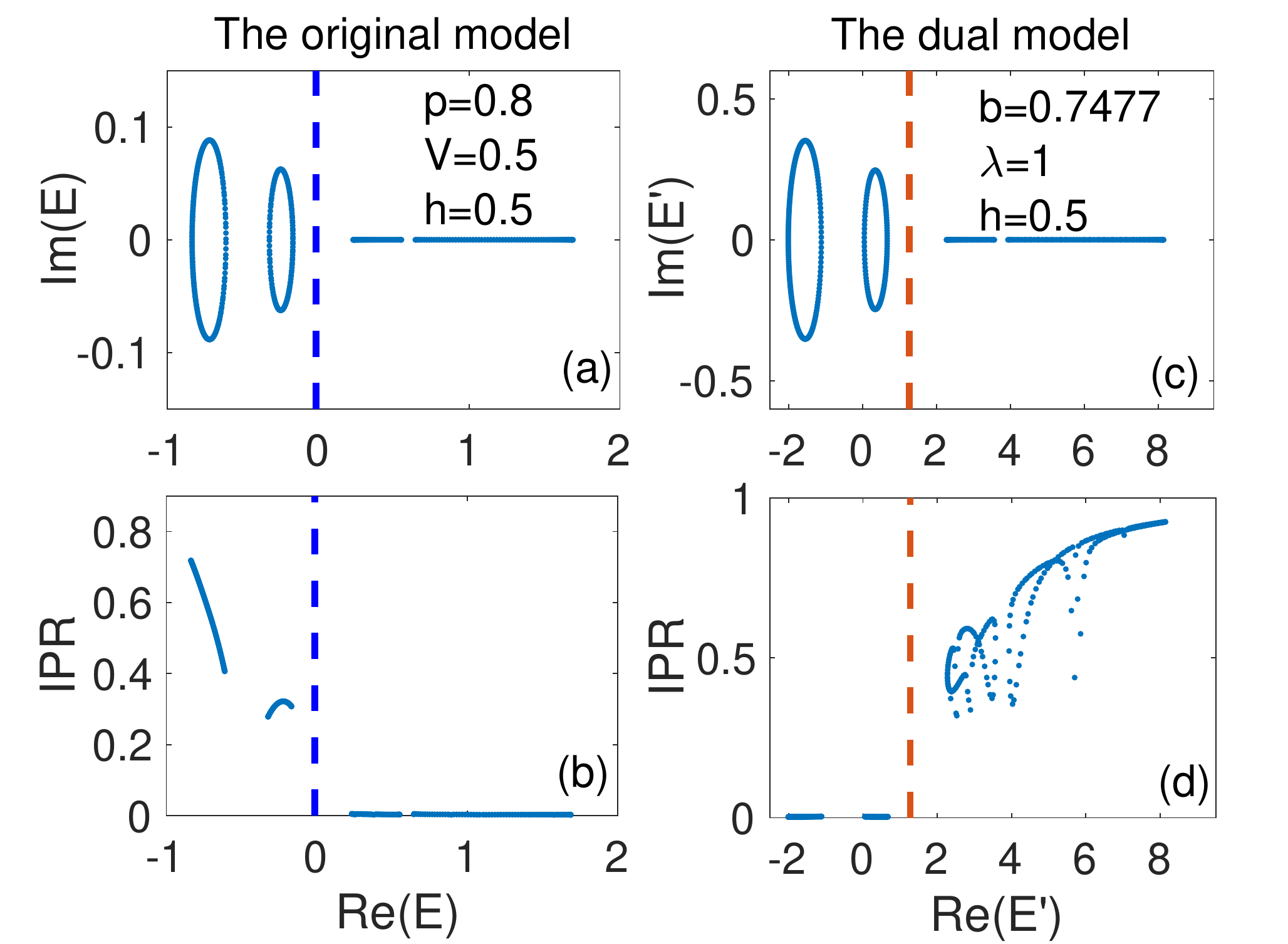}
\caption{(a) The energy spectrum and (b) the IPR of eigenstates for the model (\ref{Equation01})
with $h=0.5$, $V=0.5$, $p=0.8$ and $N=610$.
The blue dashed lines represent the exact mobility edges (\ref{MED}).
(c) The energy spectrum and (d) the IPR of eigenstates for the dual model (\ref{model})
with $h=0.5$, $\lambda=1$, $b=0.7477$ and $N=610$.
The orange dashed lines represent the exact mobility edges (\ref{LEC}).}
\label{fig2}
\end{figure}

In Fig.\ref{fig2}(a) and (b), we plot the energy spectrum and the numerical results of IPR
versus eigenenergies $E$ for the model (\ref{Equation01}) with $h=0.5$, $V=0.5$, $p=0.8$ and $N=610$.
The eigenstates of model (\ref{Equation01}) with eigenenergies $Re(E)<E_{c}$,
where $E_{c}=V \cosh(p+|h|)-1$, are localized states,
and the corresponding IPRs of these states take finite values. On the other hand,
IPRs of states with $E>E_{c}$ approach zero, corresponding to extended states.
The mobility edge $E_c$ not only separates the localized and delocalized states,
but also real and complex eigenenergies of the model.
%This clearly shows that the transition from extended to localized states and $\mathcal{PT}$-symmetry breaking transition have the same boundary.
Fig.\ref{fig2} (c) displays the energy spectrum of the dual model (\ref{model}) with the same parameters as in Fig.\ref{fig2} (a).
Energy spectrum displayed in Figs.\ref{fig2}(a) and (c) have similar structure and can be mapped to each other by using the linear relation between $E$ and $E'$ given by Eq.(\ref{energy}).
Fig.\ref{fig2}(d) shows the numerical results of IPR
versus eigenenergies $E'$. Here the IPR for the $i$th eigenstate of the dual model
is defined as $\text{IPR}^{(i)}=(\sum_{n}\left\vert u_{n}^{i}\right\vert
^{4})/\left( \sum_{n}\left\vert u_{n}^{i}\right\vert ^{2}\right) ^{2}$.
The eigenergeis of extended states for the model (\ref{model}) lie in $Re(E') < E'_{c}$,
while the localized states distribute in $E'>E'_{c}$.
\begin{figure}[tbp]
\includegraphics[width=0.5\textwidth]{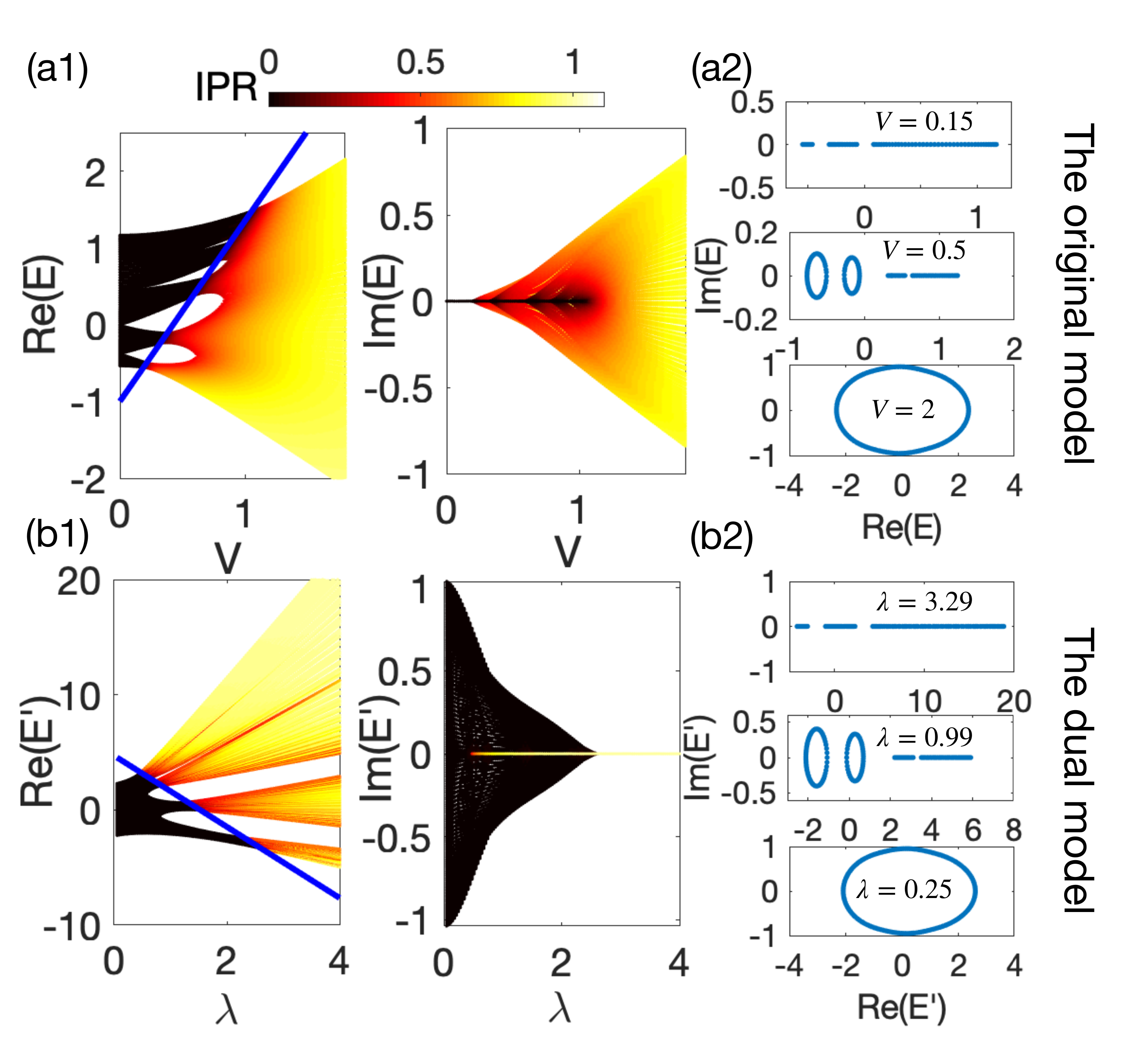}
\caption{(a1) The real and the imaginary part of the eigenvalue spectra of the original model (\ref{Equation01})
 versus $V$ for the system with $p=1$ and $h=0.5$ and $N=233$.
(a2) The energy spectrum of eigenstates with $V=0.15$, $0.5$ and $2$, respectively.
(b1) The real and the imaginary part of the eigenvalue spectra of the dual model (\ref{model})
 versus $\lambda$ for the system with $b=0.648$ and $h=0.5$ and $N=233$.
(b2) The energy spectrum of eigenstates with $\lambda=3.29$, $0.99$ and $0.25$, respectively.
The blue solid lines represent the exact mobility edges (\ref{MED}) or (\ref{LEC}).}
\label{fig3}
\end{figure}

In Fig.\ref{fig3}(a1), we plot the real and imaginary parts of eigenvalues as well as the IPR of the corresponding
wavefunctions versus the potential strength $V$ by fixing $h=0.5$ and $p=1$ for the original model (\ref{Equation01}).
The results of the dual model (\ref{model})  with corresponding parameters $h=0.5$ and $b=0.648$ are shown in Fig.\ref{fig3}(b1), where $E'$ and $\lambda$ can be mapped to $E$ and $V$ via Eqs(\ref{energy}) and (\ref{lambV}), respectively.
The blue solid lines Fig.\ref{fig3}(a1) and (b1) mark the transition points determined by Eq.(\ref{MED}) and Eq.(\ref{LEC}), respectively.
We plot spectrum structure in the complex plane with various $V$ and their corresponding $\lambda$ in Fig.\ref{fig3}(a2) and (b2), which clearly show the existence of dual relations between these two models.

We note that the model (\ref{Equation01}) has parity-time (PT)-symmetry \cite{Bender,Liu2020} and there exists a PT-symmetry unbroken region with all eigenvalues being real when $V<V_c$. On the other hand, the model (\ref{model}) has no PT-symmetry, and one may feel strange that there still exits a region with all eigenvalues being real for $\lambda> \lambda_c$.
With the help of the dual relation, we can  understand why there exists a localized region with real eigenvalues for the  model (\ref{model}). Although the model (\ref{model}) has no PT-symmetry, its dual model (\ref{Equation01}) has PT-symmetry and the extended states correspond to the real eigenvalues. It has been demonstrated that the localization transition induced by the non-Hermitian quasiperiodic potential always occurs at the PT-symmetry-breaking point \cite{Liu2020}.

\section{Summary}
In summary, we studied exact localization transition for the non-Hermitian model with exponentially decaying hopping and
complex potential, which lacks self-duality symmetry but can be mapped to a non-reciprocal Ganeshan-Pixley-Das Sarma model with only nearest-neighbor
hopping. The exact Lyapunov exponent $\gamma(E)$ of the dual model with $h=0$ can be obtained by
applying Avila's global theory. In the presence of non-reciprocal hopping, the localization-delocalization transition occurs as long as $h=\gamma(E)$, which determines analytically the mobility edges of the dual model. By using the dual transformation, the analytical expression of mobility edges for the original model is obtained. The dual transformation also constructs exact mappings between the spectra and wavefunctions of the original and dual models.

\begin{acknowledgments}
This work is supported by the National Key
Research and Development Program of China (2016YFA0300600), NSFC under Grants No.11974413 and the Strategic
Priority Research Program of CAS (XDB33000000). Zuohuan Zheng acknowledges
financial supports of the NSF of China (No. 12031020, 11671382), CAS Key Project
of Frontier Sciences (No. QYZDJ-SSW-JSC003), the Key Lab. of Random Complex
Structures and Data Sciences CAS and National Center for Mathematics and
Interdisciplinary Sciences CAS. Yongjian Wang is supported by the NSF of China (No. 12061031).
\end{acknowledgments}

\appendix

\section{The details for Lyapunov exponent}

We use Avila's global theory to calculate the Lyapunov exponent (LE) of the following Schr$\ddot{o}$dinger equation:
\begin{equation}\label{mymodel}
(Hu)_{j}=u_{j+1}+u_{j-1}+V_{j}u_{j}=E u_{j},\ j\in \mathbb{Z},
\end{equation}
where
\begin{equation}\label{mypotential}
V_{j}=2\lambda\frac{\cos(2\pi\omega j+\phi)}{1-b \cos(2\pi\omega j+\phi)},\  b \in (-1,1),\ \lambda\neq0,
\end{equation}
$V_j$ is the potential, $ b $ is the
parameter, $\phi$ is the phase, $\lambda\in\mathbb{R}$ is the coupling, and the frequency $\omega\in\mathbb{R}$ is irrational.

The Lyapunov exponent can be evaluated based on the transfer matrix technique. The model (\ref{mymodel}) can be transformed into the form
\begin{equation}
\begin{split}
\begin{pmatrix}u_{j+1}\\ u_{j}\end{pmatrix}&=M_{j}\begin{pmatrix}u_{j}\\ u_{j-1}\end{pmatrix}\\&=M_{j}M_{j-1}\cdots M_{1}\begin{pmatrix}u_{1}\\ u_{0}\end{pmatrix},
\end{split}
\end{equation}
where the transfer matrix $M_{j}$ is given by
\begin{equation}
M_{j}=\begin{pmatrix}E-V_{j}&-1\\ 1&0\end{pmatrix}.
\end{equation}
The transfer matrix is
$$T_{n}(E,\phi)=M_{n}M_{n-1}\cdots M_{1}=\prod\limits_{j=1}^{n}M_{j}.$$
The Lyapunov exponent about $T_{n}(E,\phi)$ is defined as
$$\gamma(E)=\lim_{n\to\infty}\frac{1}{2\pi n}\int \ln||T_{n}(\phi)||d\phi.$$
It is obvious that, $\gamma(E)\ge0$ due to $\text{det}T_{n}(E)=1$.

The first stage in calculation of the LE  is the complex of the phase, i.e., $\phi \rightarrow  \phi+i\epsilon)$.
The potential becomes
$$V_{j}(\phi+i\epsilon)=2\lambda\frac{\cos(2\pi\omega j+\phi+i\epsilon)}{1-b\cos(2\pi\omega j+\phi+i\epsilon)},$$
 We induce a new matrix $\widetilde{M}_{j}$, which can be written as
\begin{widetext}
\begin{equation}
\begin{split}
\widetilde{M}_{j}(\phi)&=(1- b  \cos (2\pi\omega j+\phi))M_{j}=(1- b  \cos (2\pi\omega j+\phi))
\begin{pmatrix}
E-V_{j}&-1\\
1&0
\end{pmatrix}\\
&=\begin{pmatrix}
E-( b  E+2\lambda)\cos(2\pi\omega j+\phi)&-1+ b  \cos (2\pi\omega j+\phi)\\
1- b  \cos(2\pi\omega j+\phi)&0
\end{pmatrix},
\end{split}
\end{equation}
\end{widetext}
due to the complexity of matrix $M_j$.
The transfer matrix for $\widetilde{M}_{j}(\phi)$ can be expressed as
$$\widetilde{T}_{n}(E,\phi)=\prod\limits_{j=1}^{n}\widetilde{M}_{j}(\phi)$$
The Lyapunov exponent about $\widetilde{T}_{n}(E,\phi+i\epsilon)$ is
 $$\tilde{\gamma}(E,\phi+i\epsilon)=\lim_{n\to\infty}\frac{1}{2\pi n}\int  \ln||\widetilde{T}_{n}(E,\phi+i\epsilon)|| \text{d}\phi.$$
It is easy to see that
\begin{equation}\label{relation}
\begin{split}
&\gamma(E,\epsilon)\\&=\tilde{\gamma}(E,\epsilon)-\lim_{n\rightarrow\infty}\frac{1}{2\pi n}\sum^{n}_{j=1}\ln(1- b  \cos(2\pi\omega j+i\epsilon))\\
&=\tilde{\gamma}(E,\epsilon)-\frac{1}{2\pi}\int_{0}^{2\pi} \ln(1- b  \cos(\phi+i\epsilon))d\phi\\
&=\begin{matrix}\tilde{\gamma}(E,\epsilon)-\ln\frac{1+\sqrt{1- b ^2}}{2},&if\ |\epsilon|<\ln|\frac{1+\sqrt{1- b ^2}}{ b }|.\end{matrix}
\end{split}
\end{equation}
This means that $\gamma(E,\epsilon)$ and $\tilde{\gamma}(E,\epsilon)$ has the same slope about
$\epsilon$ when $|\epsilon|<\ln|\frac{1+\sqrt{1- b ^2}}{ b }|$.

In the large-$\epsilon$ limit, we get
\begin{equation}
\widetilde{T}_{n}(E,\epsilon)=\frac{1}{2}e^{-2\pi\omega ni+|\epsilon|}\begin{pmatrix}- b  E-2\lambda& b \\- b &0\end{pmatrix}+o(1).
\end{equation}
Avila's global theory can be extended to the general case and it shows that $\tilde{\gamma}(E,\epsilon)$ is a convex,
piecewise linear function about $\epsilon\in(-\infty,\infty)$. For the model (\ref{mymodel}), in large-$\epsilon$ limit, the slope about $\epsilon$ is
always $1$, which further implies that

$$\tilde{\gamma}(E,\epsilon)=|\epsilon|+\ln f(E),$$
for large enough $\epsilon$, where $$f(E)=|\frac{ |b  E+2\lambda|+\sqrt{( b  E+2\lambda)^2-4 b ^2}}{4}|.$$

Moreover, by the convexity of $\tilde{\gamma}(E,\epsilon)$ about $\epsilon$, we have
\begin{equation}\tilde{\gamma}(E,\epsilon)\ge\ln f(E),
\end{equation}
thus
\begin{equation}\label{inequation1.11}
\gamma(E,\epsilon)\ge\max\{\ln \frac{2f(E)}{1+\sqrt{1-b^2}},0\}.
\end{equation}
Here we make use of $\gamma(E,\epsilon)\ge0$.

For the finite $0\le \epsilon<\ln|\frac{1+\sqrt{1- b ^2}}{ b }|$, by Avila's global theory, the slope of $\gamma(E,\epsilon)$
might be 1 or 0, since Lyapunov eponent $\gamma(E,\epsilon)$ is convex, moreover, the slope of $\gamma(E,\epsilon)$ in a
neighborhood of $\epsilon=0$ is nonzero if the energy $E$ is in the spectrum and the Lyapunov exponent $\gamma(E,0)>0$.
By equation (\ref{relation}), when $|\epsilon|<\ln|\frac{1+\sqrt{1- b ^2}}{ b }|$, the slope of $\tilde{\gamma}(E,\epsilon)$
is equal to that of $\gamma(E,\epsilon)$. Thus, when $\gamma(E,0)>0$, $0\le \epsilon<\ln|\frac{1+\sqrt{1- b ^2}}{ b }|$
and $E$ is in the spectrum, the slope of $\tilde{\gamma}(E,\epsilon)$ is also 1.

When $\gamma(E,0)>0$, since the Lyapunov exponent $\tilde{\gamma}(E,\epsilon)$ is convex and continuous,
thus the slope of $\tilde{\gamma}(E,\epsilon)$ is always 1, which implies that
\begin{equation}
\tilde{\gamma}(E,\epsilon)=|\epsilon|+\ln f(E),
\end{equation}
for any $\epsilon\in(-\infty,\infty)$. According to the equation (\ref{relation}) and the non-negativity of Lyapunov exponent $\gamma(E,\epsilon)$, we have
\begin{equation}\label{equation1.13}
\gamma(E,0)=\max\{\ln \frac{2f(E)}{1+\sqrt{1-b^2}},0\},
\end{equation}
if $\gamma(E,0)>0$ and $E$ is in the spectrum.

When $\gamma(E,0)=0$, the right side of the equation (\ref{inequation1.11}) is 0.
Thus the energy $E$ also satisfies the equation (\ref{equation1.13}).
Based on above discussion and analysis, we deduce
\begin{equation}\label{lyapunov}
\gamma(E,0)=\max\{\ln \frac{2f(E)}{1+\sqrt{1-b^2}},0\},
\end{equation}
for every $E$ in the spectrum.
\\
\\
\section{Mobility edges}
The mobility edge can be roughly determined by $\gamma_c (E)=0$,
which gives
$$|bE+2\lambda|=2.$$
The more accurate mobility edge can be obtained by operator theory.

The operator theory tell us that the spectrum $\{E_j\}$ of model (\ref{mymodel}) and $V_{j}$ have
the relation $\{E_j\}\subseteq[-2+\min\limits_{j}(V_{j}),2+\max\limits_{j}(V_{j})]$. Thus, when $\lambda>0$,
we have
\begin{equation}\label{cond1}
\{E_j\}\subseteq[-2-\frac{2\lambda}{1+b},2+\frac{2\lambda}{1-b}],
\end{equation}
and when $\lambda<0$,
we have
\begin{equation}\label{cond2}
\{E_j\}\subseteq[-2+\frac{2\lambda}{1-b},2-\frac{2\lambda}{1+b}].
\end{equation}
The eigenenergies of localized states satisfy $\gamma(E)=\gamma(E,0)>0$,
which can give us that
$$|bE+2\lambda|>2.$$
 First, we consider the case $\lambda>0$.
Assume $bE+2\lambda<-2$, i.e.
\begin{eqnarray}\label{SS1}
\left\{\begin{matrix}
E<\frac{-2-2\lambda}{b},&b>0,\\
E>\frac{-2-2\lambda}{b},&b<0.
\end{matrix}\right.
\end{eqnarray}
From condition (\ref{cond1}), we can get
\begin{eqnarray*}
-2-\frac{2\lambda}{1-b}\leq E \leq2+\frac{2\lambda}{1-b}.
\end{eqnarray*}
Since
\begin{eqnarray*}
\frac{-2-2\lambda}{b}<-2-\frac{2\lambda}{1+b},
\end{eqnarray*}
with $b>0$ and
\begin{eqnarray*}
\frac{-2-2\lambda}{b}>2+\frac{2\lambda}{1-b},
\end{eqnarray*}
with $b<0$,
this leads to contradictions.  When $\lambda>0$, the eigenenergies of localized states
only satisfy $b E+2\lambda>2$.

Secondly, we consider the case $\lambda<0$.
Assume $b E+2\lambda>2$, i.e.
\begin{eqnarray}\label{SS2}
\left\{\begin{matrix}
E>\frac{2-2\lambda}{b},&b>0,\\
E<\frac{2-2\lambda}{b},&b<0.
\end{matrix}\right.
\end{eqnarray}
From condition (\ref{cond2}), we can get
\begin{eqnarray*}
-2+\frac{2\lambda}{1-b} \leq E \leq 2-\frac{2\lambda}{1-b}.
\end{eqnarray*}
Since
\begin{eqnarray*}
\frac{2-2\lambda}{b}>2-\frac{2\lambda}{1+b}
\end{eqnarray*}
with $b>0$,
and
\begin{eqnarray*}
\frac{2-2\lambda}{b}<-2+\frac{2\lambda}{1-b}
\end{eqnarray*}
with $b<0$,
this leads to contradictions.  When $\lambda<0$, the eigenenergies of localized states
only satisfy $b E+2\lambda<-2$.

In conclusion, for the eigenenergies of localized states, the Lyapunov exponent $\gamma(E)>0$ if and only if
$$\left\{\begin{matrix}
b E>2\text{sgn}(\lambda)(1-|\lambda|),&\lambda>0,\\
b E<2\text{sgn}(\lambda)(1-|\lambda|),&\lambda<0,
\end{matrix}\right.$$
it is equal to
$$\text{sgn}(\lambda)bE>2(1-|\lambda|).$$
Thus, for the eigenenergies of extended states with Lyapunov exponent $\gamma(E)=0$, we can get
$$\text{sgn}(\lambda)bE<2(1-|\lambda|).$$
The mobility edge is $$bE=2\text{sgn}(\lambda)(1-|\lambda|).$$

\end{document}